\documentstyle[11pt]{article}
\begin{document}

\begin{flushright}
BRX TH-529 \end{flushright}

\vspace*{1in}

\begin{center}
{\large\bf ``String Theory:}

{\large\bf a theory in search of an experiment"}

\vspace{.3in}

Howard J. Schnitzer\footnote{email: schnitzr@brandeis.edu.
~~Supported in part by the DOE under grant DR-FG02-92ER40706.}\\
Martin A. Fisher School of Physics\\ Brandeis University\\
Waltham, MA 02454

\vspace{.5in}

This essay was presented at a celebration for\\
Professor S.S.\
Schweber on October 8, 2003\\
at the Dibner Institute, MIT\\
\end{center}

\vspace{.5in}

\begin{abstract}
\noindent It is argued that string theory may pose new conceptual
issues for the history and philosophy of science.
\end{abstract}

\newpage

I've been in particle physics since 1954, so I've had the good
fortune to witness the emergence of that subject from nuclear
physics, and have participated in its development, right through
its present incarnation.  I came to Brandeis University in 1961,
so that for almost my entire career in particle theory, I've had
the privilege to discuss and debate a wide variety of issues
relating to the development of particle physics, and their
interpretation in a historical context with Sam.  In that spirit,
I offer this discussion of string theory to the celebration of
Sam's accomplishments.

I have been a practicing string theorist since the mid 1980's, so
my perspective is as one immersed in the subject.  In my opinion
string theory raises new issues for the history of physics which
need be considered and clarified.  Others are more competent than
I am to comment on deeper historical and philosophical
implications relevant to the subject, so I leave interpretations
to them.  My role will be to point out issues that I believe need
further analysis.  As yet, string theory's place in the history of
physics is uncertain; at best it is history in the making.  If it
succeeds in providing a description of nature it will be one more
example of a ``top down" theory, {\it i.e.}, one which is driven
by theoretical issues, and not by experiment, at least not in its
formative stages.  That is, it draws the attention of workers to
the field because of the elegance and consistency of its
formulation and not as a response to experiments which call out
for explanation \cite{marolf}.  There are well recognized
``top-down" theories in our subject; 1) general relativity; 2) the
Dirac theory of the electron; and 3) the non-abelian gauge
theories, ({\it i.e.}, Yang-Mills theory) are prime examples. Each
of these were also theories in search of experiments, and they are
with us today because eventual confirming experimental results
became available.

In each of these three examples, crucial experiments followed
within a reasonable period of time after the presentation of the
theory.  General relativity was formulated by Einstein so as to
respect general covariance and the equivalence principle as
underlying features. The residual precesion of the perihelion of
Mercury and the deflection of light of stars seen during solar
eclipses convinced many soon after that general relativity had
something to say about nature. Applications to cosmology and
astrophysics are now standard.

The theory of the electron was formulated by Dirac in  order to
combine special relativity with quantum mechanics.  The discovery
of the positron, and an understanding of the magnetic moment of
the electron provided confirmation of the Dirac equation, and made
it a recognized tool of the theorist.  The program of string
theory to make general relativity compatible with quantum field
theory, is in some sense a logical descendant of the Einstein and
Dirac theories.  By contrast, string theory is still in the
process of development, without as yet any experimental
confirmation of the underlying ideas, let alone detailed
experimental predictions.  Nevertheless, the issues addressed by
string theory are very compelling, given the antecedents of
Einstein and Dirac.

Of course there have been a large number of ``bottom-up" theories
in particle physics as well.  These developed from an experiment
or series of experimental results which require explanation.  A
prime example in particle physics is provided by the proliferation
of ``elementary" particles in the 1960's.  This prompted several
competing theories of the strong interaction symmetries, with
Gell-Mann's SU(3), {\it i.e.}, the eight-fold way, being the
winner of the competition.  The crucial experiment in this program
was the discovery of the $\Omega^-$ baryon, which provided
convincing evidence for the Gell-Mann scheme.  The desire for a
more fundamental description of the SU(3) of strong-interactions
gave rise to the Gell-Mann/Zweig quark model, which in this
context should also be considered a bottom-up theory.

Those of us who are active in string theory hope that it  will be
one more example of a successful top-down theory, but we are not
there yet.  We are still hoping for any experiment or a series of
experiments to support the underlying ideas of this class of
theories.  (More about this later.)  Why do many of us consider
string theory so appealing?  The short answer is that it is the
only known consistent theory which combines quantum mechanics and
gravity.  But there is more.  Perhaps it will be a theory that
subsumes the standard model of particle physics, and leads us to
explanation of new phenomena?  Hopefully time will tell.

A brief historical perspective about string theory \cite{marolf}
and major changes of direction motivated by crucial calculations
or new theoretical insights may place the discussion in
perspective. Order emerged in the study of the strong-interactions
with Gell-Mann's SU(3) organizing hadron multiplets into
representations of the symmetry group SU(3). The Gell-Mann-Okubo
formula gave a precise prediction for the symmetry breaking
patterns, which were indeed observed.  Other regularities were
also observed. Resonances could be organized into a
(Chew-Frautchi) plot of angular momentum vs. (mass)$^2$, {\it
i.e.}, $J$ vs. $m^2$, for particles and resonances.  Remarkably
these are a series of parallel straight lines, so-called Regge
trajectories.  It was observed that scattering of hadrons at high
energies could be described in terms of the exchange of Regge
trajectories between the colliding particles, equivalent to the
exchange of particles of arbitrary high spin and correlated mass
between the scatterers. This is a remarkable extension of the
Yukawa idea of meson exchange between protons.  In fact in a
two-body scattering process, schematically written 1+2
$\rightarrow$ 3+4 (s-channel) there are two other associated
processes 1+$\bar 4$ $\rightarrow$ $\bar 2$+4 (t-channel) and
1+$\bar 4$ $\rightarrow$ $\bar 2$+3  (u-channel). The observation
was that all three processes were related, and the high-energy
cross-section for the set of three-processes was given by a single
amplitude.  That is, the three processes were related by so-called
``crossing symmetry". The empirical result was that this single
master amplitude could be described by Regge exchange in just one
of the three scattering channels, contrary to what is expected
from Feynman diagrams. This feature of high-energy scattering
processes was termed ``hadron duality", where the Regge-exchange
description of the scattering processes need be considered in only
one of s, t, or u channels. Not all three.  The first attempt at
explanation was provided by an {\it ad hoc} formula due to
Veneziano \cite{veneziano}. Subsequently Nambu \cite{nambu} and
Goto independently noted that the results of the Veneziano model
and its generalization to multiparticle processes (the
dual-resonance model) could be described by a theory of
relativistic vibrating strings.

This string theory of strong interactions was soon set aside for a
variety of reasons.  For example it disagreed with scattering at
high energies and large momentum transfers. [The original
motivation for hadron duality involved small momentum transfers.]
The data indicated that these large momentum transfer processes
were much harder than the soft cross-sections predicted by the
dual resonance models.  That is, cross-sections decreased as a
power-law in the large momentum transfer, measured at fixed
energy, rather than the predicted exponential fall-off.  Further,
deep-inelastic scattering gave evidence for quarks, which did not
fit into the dual resonance model in a natural way.

These results indicated that the dual resonance model, or the
equivalent string model, was not a viable theory of the strong
interactions.  In addition, the dual resonance model predicted the
existence of a massless state with spin $J$=2, which had no role
in the strong interactions.  Joel Scherk and John Schwarz
\cite{scherk} turned this latter difficulty into an advantage,
proposing that the massless, spin 2 state was the graviton, and
that the Nambu-Goto string theory was in fact a theory of
gravitational interactions. The modern theory of strings was
launched.

Thus, the Nambu-Goto string theory was reinterpreted as a
fundamental string, with the Planck mass replacing the hadron
mass-scale of (1Gev).  In the new understanding, the slope of
Regge trajectories was $\alpha^\prime$ = ($\Delta J$=1)/(Planck
mass)$^2$, and not ($\Delta J$=1)/(1 Gev)$^2$.  So what had begun
as a phenomenological theory of hadrons, reemerged as a theory of
gravity, coupled to particle states.  Various quantum field
theories coupled to gravity could be obtained in well-controlled
low-energy limits. To restate the situation, what was originally
an attempt at a bottom-up theory of hadrons, had a new incarnation
as a top- down theory entirely motivated by theoretical concerns.
It was not yet clear whether this theory was internally
consistent, let alone a description of nature.  Various versions
of string theory were found, but there were consistency issues. In
order to be consistent with Lorentz invariance and unitarity
(conservation of probability) bosonic strings (fermions absent)
had to be formulated in 26-dimensions, with 22 of these compact if
contact with our 4-dimensional world was to be achieved.  It is
presumed that the compact dimensions are much too small to be
observed. [Planck size.]  However, this theory had an unpleasant
feature. It contains a tachyon state, {\it i.e.}, one with $m^2
<0$, which ruled it out for serious consideration, except as a
possible practice field. In order to have a completely consistent
string theory, one had to add in the formulation supersymmetry, a
symmetry which exchanges bosons $\leftrightarrow$ fermions.  Then,
in order to have Lorentz invariance, unitarity, and absence of
tachyons, one was restricted to D=10, of which 6 of these are
extremely small, and compact, if contact with our D=4 world was to
be made \cite{marolf}.

However, other theoretical difficulties remained.  A number of
string theories exhibited ``anomalies".  Anomalies are good
symmetries of the classical theory, which are broken by the
quantum theory, in a way that rendered the theories inconsistent.
In string theory, the difficulties occurred with gravitational
anomalies.  In the 1984 string revolution Michael Green and John
Schwarz \cite{green} found 3 distinct consistent string theories
in D=10, all of which were supersymmetric.  [Subsequently 2 more
(heterotic) strings were found by Gross, {\it et al.}
\cite{gross}, to establish the 5 only consistent string theories
in 10-D.] The crucial calculations of Green and Schwarz were so
tight in their logic and compelling in their conclusions that they
convinced a number of workers (including me) that this theory was
extremely deep and possibly had something to do with nature. It
was certainly the first and only theory which consistently
combined general relativity with quantum field theory.  As a
bonus, each of the consistent string theories are ultraviolet
finite, in that the ultraviolet divergences which infest the usual
quantum field theories are absent.  No renormalization is needed.
In the euphoria of the moment, several workers proclaimed that one
was on the road to a theory of everything. This overselling of the
beautiful achievement of combining general relativity with quantum
theory certainly created antagonism to the program, much of which
seems to have dissipated.

An ongoing difficulty with any supersymmetric theory is that
supersymmetry is at best a broken symmetry.  We have yet to find
bose $\leftrightarrow$ fermi partners for the known particles.  If
found, the partners would not have the same masses, as the
symmetry will be a broken one, with characteristic mass
differences of at least 1 Tev.  It turns out that breaking
supersymmetry in string theory seems inevitably to lead to a
cosmological constant in the gravitation sector.  Workers  ({\it
e.g.}, T.\ Banks \cite{banks}) are presently trying to turn this
feature into an advantage, given the observed cosmological
constant found from analyses of cosmic microwave background
radiation.  The attempt to correlate the observed cosmological
constant with a prediction of the mass-scale characterizing
supersymmetry breaking is work in progress.

A deep insight; that of a chain of string dualities was discovered
by Chris Hull, Paul Townsend, Edward Witten \cite{hull} and
others. This provided a demonstration that the 5 consistent string
theories in 10-D were in fact different manifestations of a single
theory. These 5 theories are linked by dualities to a 6th theory
in 11-D, whose low-energy realization is 11-D supergravity.
[Supergravity was an initial attempt to combine supersymmetry and
quantum field theory with gravity, but it was shown early on at
Brandeis that it too was nonrenormalizable \cite{deser}, with an
infinite number of undetermined constants. Supergravity is now
considered to be the low-energy limit of string theories.]  A
puzzle had been that there were 6 natural supergravity theories in
D$\geq$10 [5 in 10-D and 1 in 11-D], but there were only five 10-D
string theories.  It has been proposed that there is an 11-D
master theory, M-theory \cite{townsend}, which subsumes all 6
incarnations, and which links these 6 by dualities, but M-theory
still lacks a complete formulation.

Let me sketch some aspects of the chain of string dualities that
links these string theories, to give a flavor of the idea.
Consider a particular string theory in 10-D, {\it before} any of
the dimensions have been compactified to reach D=4.  Then let the
10th dimension be compactified on a circle of radius $R$.  Call
this theory IIA.  There is another theory compactified on a radius
1/$R$. [I use dimensionless units in units of Planck length.] Call
this theory IIB.  It turns out that the IIA theory compactified on
radius $R$ is {\it identical} to IIB theory compactified at radius
1/$R$. [Parenthetically, this leads to a string uncertainty
principle.  A new conceptual idea.  One cannot probe distances
shorter than the Planck length in a scattering process, as a
duality $R\rightarrow 1/R$ maps this to a distance scale {\it
larger} than the Planck length.] There is another duality,
S-duality, which relates one theory with string coupling constant
$g_s$ to another string theory with coupling constant 1/$g_s$.
This is the string analogue of the electromagnetic duality of
Maxwell's theory under which $E\rightarrow B$ and $B\rightarrow
-E$.  This gives a glimpse of the chain of dualities, which links
all 5 string theories, as well as the 11-D supergravity. Hence,
the idea that there is just one master theory in 11-D, M-theory,
which however is still not completely characterized.

Recall that to make any possible contact with the real world, we
must compactify 6 of the 10-dimensions, leaving 4-D Minkowski
space, in the low-energy limit.  There are a large number of ways
of carrying out this compactification, each leading to potentially
different predictions of 4-dimensional physics.  With abuse of
language, we call these different compactifications different
theories.

A recent deep insight was the principle of  holography ('t~Hooft;
Susskind; Maldacena) \cite{thooft}. It was proposed, supported by
many concrete examples that verified the idea, that a consistent
quantum theory of gravity in (D+1) compact bulk dimensions [string
theory is the only known example] is dual to a quantum field
theory without gravity living on the D-dimensional boundary.  This
duality in detail implies that correlators of the bulk and
boundary theories are related. The principle also states that
there is at most (one degree of freedom)/Planck area on the
boundary.  This point of view has led to a description of
black-hole thermodynamics in terms of string microstates, for
certain well-chosen examples \cite{strominger}. Thus,
string-theory seems to provide the statistical mechanics which
underlie black-hole thermodynamics.  Optimists hope that further
developments in this direction will solve the black hole
information paradox.  Loosely speaking, throw the encyclopedia
into a black-hole, and out comes uncorrelated thermal Hawking
radiation.  Where has the information gone?  But more needs to be
done to see if the paradox can be resolved. Holography has also
provided many new insights into the strong-interactions of quantum
field theories.  This too remains an active area of investigation.

All these beautiful developments are dominated by theoretical
considerations, but not by choice.  What about prospects for a
crucial experiment?  In the next generation of experiments, which
will be carried out at the LHC (CERN) now under construction,
there will be dedicated searches for (broken) bose-fermi symmetry,
{\it i.e.}, supersymmetry.  If found, this will be good news for
string theories, as supersymmetry seems to be a necessary, but not
sufficient, ingredient to confirm string theory.

Another issue being actively explored is the possibility that not
all of the 6 hidden dimensions are small ({\it i.e.}, of Planck
size).  Some might be considerably larger (mm?) \cite{arkani}.
Considerable effort is going into formulating experiments to test
for large compact dimensions.  [Extra dimensions is a natural
descendent of the top-down model of Kaluza-Klein.]  Their
experimental discovery would be sufficient to confirm this
essential idea of string theory, without pointing to any specific
version of the theory. So, discovery of large extra dimensions is
sufficient to support one of the fundamental features of string
theory, but is not necessary.  It could be that all hidden
dimensions are of Planck size.

So supersymmetry and/or large extra dimensions would give support
to the general ideas of string theory, but would not provide the
crucial experiments in the historical sense. Can we foresee a
crucial experiment?  Not at the present time.  String theorists
are presently making excursions into cosmology, so perhaps there
will be a distinctive signature in that application of the theory.
Perhaps there is a particularly attractive compactification of 10
$\rightarrow$ 4 dimensions which agrees very well with the
standard model, fixing some or all of the free parameters, and
makes distinctive predictions for future experiments.  But this
seems to be a very distant prospect.

There is another unpleasant possibility that is presently being
discussed.  Suppose that there are a large number of different
string theory compactifications (1000 say, to as many as
$10^{100}$ claimed by Douglas \cite{douglas}), which make
predictions that closely resemble those of the standard model and
its extensions within experimental error. [It should be emphasized
that there are no known examples, since we don't have a single
example of this class!] Instead, what is studied are the
statistics of the number of string theories which make low-energy
predictions close to each other. Perhaps these overlapping
theories may be distinguished by cosmology.  If not we are faced
with a deep and disturbing problem.  We would have a large number
of theories which agree with any conceivable experiment, but lack
a crucial experiment to distinguish them. Simplicity does not seem
to favor any particular theory of this class, so this is
apparently not a decisive criterion.  How is one to assess such a
class of theories? This is certainly science and not mathematics,
since they would correctly predict all possible accessible
experiments, all within experimental accuracy.  If indeed this is
what lies in the future of string theory, in my opinion it opens
new issues in the history of physics, ripe for new
interpretations.

Perhaps this point of view is too pessimistic, and represents a
``pendulum-like" swing away from the early euphoria that claimed
that string theory was a theory of everything. The reaction seems
to have swung  too far in the opposite direction, as not a single
concrete example exists to illustrate the concerns of the
pessimists.  They are certainly considering logical possibilities,
but we just don't know how the story will evolve.

In any case, string theory is still work in progress, with final
judgment as to its role in physics still to be decided, and there
may be further string revolutions before the ultimate form of the
theory takes shape.  As I've attempted to indicate, this
particular ``top-down" theory raises a number of problems in the
assessment of the status of a theory, and its confirmation by
experiment.  Hopefully we will see theoretical and conceptual
clarification in the future.

\end{document}